\documentclass[reqno]{amsart}
\usepackage{graphicx,epsfig}
\usepackage{epstopdf}
\usepackage{enumitem}
\usepackage{todonotes}
\usepackage[numbers,sort&compress]{natbib}
\usepackage{hyperref}
\usepackage{soul}
\usepackage[foot]{amsaddr}
\usepackage[final,color,notref,notcite]{showkeys}

\usepackage[charter]{mathdesign}

\definecolor{labelkey}{rgb}{0,.56,.7}

\presetkeys{todonotes}{color=black}{}

\DeclareMathAlphabet{\pazocal}{OMS}{zplm}{m}{n}   %  mathcal

\newcommand{\Ncal}{\pazocal{N}}

\newcommand*{\at}{@}

\newcommand{\nn}{\nonumber}
\def\wh{\widehat}
\def\dg{\dagger}
\def\df{\overset{\mathrm{df}}{=}}
\newcommand{\ket}[1]{\mathop{|#1\rangle}\nolimits}

\newcommand{\kbr}[2]{| #1\rangle\!\langle #2 |}

\newcommand{\Tr}[1]{\mathop{{\mathrm{Tr}}_{#1}}}
\newcommand{\id}{\mathop{{\mathrm{id}}}\nolimits}
\newcommand{\rank}{\mathop{{\mathrm{rank}}}\nolimits}

\def\supp{\mathop{\mathrm{supp}}}
\def\hmineps{H^\ve_{\mathrm{min}}}
\def\hmaxeps{H^\ve_{\mathrm{max}}}
\def\hmin{H_{\mathrm{min}}}
\def\hmax{H_{\mathrm{max}}}
\newcommand{\dif}[1]{\mathrm{\,d} #1}             % dx
\newcommand{\diff}[1]{\mathrm{\,d}^2 #1}             % d^2x

\def\a{\alpha}
\def\b{\beta}
\def\g{\gamma}

\def\e{\epsilon}
\def\ve{\varepsilon}
\def\vr{\varrho}

\def\ka{\kappa}
\def\om{\omega}

\def\s{\sigma}

\def\la{\lambda}

\def\vt{\vartheta}

\def\om{\omega}

\def\bbR{\mathbb{R}}

\sodef\so{}{.065em}{.4em plus1em}{2em plus.1em minus.1em}

\usepackage{color}

\makeatletter
\def\@setemails{%
\ifnum\theg@author > 1
\mbox{\hspace{-4mm}{\itshape E-mail}:\space}{\ttfamily\emails}.
\else
\mbox{\hspace{-4mm}{\itshape E-mail}:\space}{\ttfamily\emails}.
\fi%
}
\makeatother

\makeatletter
\def\@setfoot@addresses{
\def\author##1{\hspace{-4mm}\setlength{\parindent}{0pt}}%
\def\\{\unskip, \ignorespaces}%
\newif\if@firstaddr
\@firstaddrtrue
\def\address##1##2{%
\if@firstaddr\@firstaddrfalse\else\par\fi
\@ifnotempty{##1}{(\ignorespaces##1\unskip) }%
{\scshape\ignorespaces##2}%
}%
\def\email##1##2{}%
\def\curraddr##1##2{}%
\def\urladdr##1##2{}%
\addresses
}
\makeatother

\makeatletter
\def\@setkeywords{%
  {\itshape\hspace{-4.8mm} \keywordsname.}\enspace \@keywords\@addpunct.}
\makeatother

\begin{document}

\title[Finite-key security analysis for multilevel quantum key distribution]{\hspace*{1.4cm}\so{  Finite-key security analysis for multilevel  \\   quantum key distribution}}

\begin{abstract}
We present a detailed security analysis of a $d$-dimensional quantum key distribution protocol based on two and three mutually unbiased bases (MUBs) both in an asymptotic and finite key length scenario. The finite secret key rates (in bits per detected photon) are calculated as a function of the length of the sifted key by (i) generalizing the uncertainly relation-based insight from BB84 to any $d$-level 2-MUB QKD protocol and (ii) by adopting recent advances in the second-order asymptotics for finite block length quantum coding (for both $d$-level 2- and 3-MUB QKD protocols). Since the finite and asymptotic secret key rates increase with $d$ and the number of MUBs (together with the tolerable threshold) such QKD schemes could in principle offer an important advantage over BB84. We discuss the possibility of an experimental realization of the 3-MUB QKD protocol with the orbital angular momentum degrees of freedom of photons.
\end{abstract}

\author{Kamil Br\'adler}

\email{kbradler\at uottawa.ca}

\address[Kamil Br\'adler, Anne Broadbent]{Department of Mathematics and Statistics, University of Ottawa, Canada}

\address[Kamil Br\'adler]{Max Planck Centre for Extreme and Quantum Photonics, University of Ottawa, Canada}

\author{Mohammad Mirhosseini}

\address[Mohammad Mirhosseini]{The Institute of Optics, University of Rochester,
New York, 14627, USA}

\author{Robert Fickler}

\address[Robert Fickler]{Department of Physics and Max Planck Centre for Extreme and Quantum Photonics, University of Ottawa, Ottawa, K1N 6N5, Canada}

\author{Anne Broadbent}

\author{Robert Boyd}
\address[Robert Boyd]{Department of Physics and  Max Planck Centre for Extreme and Quantum Photonics, University of Ottawa, Ottawa, K1N 6N5, Canada, The Institute of Optics, University of Rochester, Rochester,
New York, 14627, USA}

\maketitle

\thispagestyle{empty}
\allowdisplaybreaks

\section{Introduction}

It has been more than 30 years since the proposal of the first quantum key distribution (QKD) protocol -- BB84~\cite{BB84}. The ultimate goal of a QKD protocol is to establish a secure key between two parties for a further cryptographic use; in this context, quantum mechanics is a powerful ally of the legitimate parties. Therefore, it is advantageous to generate the key by distributing and measuring quantum states. Contrary to communication with classical signals, for quantum states there exists a fundamental trade-off between how much information a classical or quantum adversary can get and how much the quantum system is disturbed. For example, the most straightforward strategy of simply copying a quantum state does not work~\cite{Wootters:1982ex, Milonni:1982es}. A significant amount of effort has been invested in proving the security of BB84 and subsequent QKD protocols (starting with its proper definition~\cite{devetak2005private,rennersecurity}) and experimental realization~\cite{scarani2009security}.

Most of the modern QKD schemes rely on two-level quantum systems (qubits) as  quantum information carriers. This is especially easy to achieve using the photon polarization degree of freedom. The theoretical background as well as the experimental techniques are mature.  However, quantum $d$-level states (qudits) have attracted much attention recently because they naturally offer higher quantum information transmission rates and together with continuous variables are promising candidates for next generation quantum information processing. In this approach, the information is encoded onto $d$ distinct orthogonal states, for which in principle there is no upper limit on~$d$. In the context of QKD, the $d$-level  protocols not only offer a great potential to increase the transmitted key rate but they are also known to be more resilient to errors~\cite{Cerf:2002fp}. Experimentally, high-dimensional quantum states have been realized as discrete time-bins~\cite{AliKhan:2007ia}, positions~\cite{Walborn:2006jv} or angular momenta~\cite{Mirhosseini:2015fy} in lab-scale proof-of-principle tests. They have also been successfully studied under real world environmental conditions where air turbulence or inter-modal coupling in fibers have to be taken into account \cite{Malik:2012ka,Rodenburg:2014dn}.

The experimental efforts for realization of multidimensional QKD has primarily relied on employing two mutually unbiased bases (MUBs).  However, it is known that using only two MUBs for $d=2$ does not realize the full potential of a qubit-based QKD. Instead, by using three MUBs we are rewarded by an increase in the maximum tolerable error rate in a QKD protocol known as the six-state protocol~\cite{bruss1998optimal}.  Considering this observation, it is expected that using more than two MUBs would provide enhancement in the security of the $d$-dimensional QKD protocols. It is well known that for $d$ a prime number or the power of a prime, the maximum number of MUBs in a $d$-dimensional Hilbert space is $d+1$~\cite{Wootters:1989ba,bandyopadhyay2002new}. For the non-prime dimensions, the number of MUBs is a major open problem. However, it is perhaps less well known that there always exists three MUBs for any $d$~\cite{durt2010mutually}. Motivated by this fact, we present a comprehensive security analysis for $d$-level QKD with two and three MUBs. Our main contribution in this paper is the calculation of the secret key rate upper bounds for discrete $d$-dimensional QKD protocols using two and three MUBs. We exemplify the key rate calculations on $d=2$ to $7$ but our approach can be immediately applied for any $d$. The secret key rates are calculated in both the asymptotic and finite key length scenario. In the asymptotic case, the 2-MUB rates reproduce the previously known results~\cite{bechmann2000quantum,bruss2002optimal,Cerf:2002fp,scarani2008quantum,scarani2009security,abruzzo2011quantum,bratzik2011min,cai2009finite,renner2005information,ferenczi2012symmetries,sheridan2010security} but to our best knowledge the analytical results we obtain for 3-MUB rates and  for any $d$ are novel and the corresponding adversarial channels haven't been studied before (only the $d=2$ case reduces to the well studied six-state protocol~\cite{bruss1998optimal}). The main reason  to reproduce the already known results for the 2-MUB QKD protocol is the calculation method that may not be familiar to the practitioners of QKD. It can be summarized as ``ab initio'' since our starting point is the private classical capacity and the quantum capacity of a quantum channel~\cite{devetak2005private} and we systematically derive the well-known expressions for the secret key rate. The main result of the asymptotic part of our analysis is the secret key rate calculation for the 3-MUB protocol and the derivation of the tolerable threshold for the error rate. We found that the threshold quite substantially increases accompanied by the increase of the secret key rate\footnote{The secret key rate units are bits per channel where the channel is understood as a completely positive map whose exact form will be derived. Therefore our notion of a channel differs from its typical use in quantum optics experiments. A quantum channel is said to be realized in the QKD context whenever the photon is detected and used in the process of secret key extraction (not discarded). Knowing the number of realizations of the channel per second gives us the total number of secret bits per a unit of time, sometimes perhaps confusingly also called a rate.} as envisaged by the comparison of BB84 and the six-state protocol. Our results justify the overlapping numerical results presented in~\cite{coles2015unstructured}.

The second part of our analysis is the study of QKD in the non-asymptotic regime of a finite number of exchanged signals. We follow two different routes leading to excellent (achievable~\cite{privComRenner}) upper bounds on the secret key rates even for a relatively low number of signals. The first approach is the generalization of the uncertainty relation-based approach pioneered in~\cite{tomamichel2012tight} for two MUBs and $d=2$. We generalize the key step spelled out in~\cite{tomamichel2015rigorous} for any $d$ and using the large deviation estimate for the number of errors in the non-sacrificed part of the sifted key we derived the corresponding secret key rates. The intermediate step includes a numerical optimization over the ratio of dits in the secret key rates that are sacrificed for the parameter estimation purposes. As the number of sifted bits asymptotically increases the portion of sacrificed bits tends to zero~\cite{lo2005efficient} and the secret key rates approach the asymptotic ones derived previously. For another approach to the non-asymptotic regime see~\cite{hayashi2012concise,hayashi2015quantum}.

The uncertainty-relation-based method is, however, not known to be applicable to the 3-MUB QKD protocol~\cite{tomamichel2012tight}. More precisely, it can be enforced even for three MUBs but our attempts lead to awfully suboptimal rates. Hence we adopt a different strategy. Using the recent advances in the second-order asymptotics for the quantum coding rates~\cite{tomamichel2013hierarchy} we use the expansion of the relevant entropic quantity (the smooth min-entropy) in terms of the conditional entropy variance~\cite{li2014second,tomamichel2013hierarchy} and expand the  decoupling exponent of what is essentially a  one-shot decoupling lemma~\cite{tomamichel2012tight}. The resulting rates are calculated both in the 2- and 3-MUB QKD scenario. In the latter, the resulting secret key rates are better for any $d$ compared to the basic estimate first brought by Renner in~\cite{renner2005information} that is used as a template in almost all finite key studies. Since the 3-MUB QKD protocol for any $d$ seems to be systematically studied for the first time here, it therefore establishes the best known secret key rates. The second-order asymptotic expansion also beats Renner's rates for the 2-MUB QKD protocol (for any $d$) but it is not as good as the uncertainty-relation-based estimates. This is the expected kind of behavior.

The remainder of the paper is structured as follows. In Sec.~\ref{sec:prelim} we introduce the minimal background material and notation for our approach to calculate the asymptotic secret key rates and collect several rudimentary facts about the Pauli group for qudits and mutually unbiased basis. We also recall the Choi-Jamio\l kowski state-map correspondence. The asymptotic rates for 2- and 3-MUB QKD protocol are calculated in Sec.~\ref{sec:QKDasymptotic}. In Sec.~\ref{sec:finiteSizeKey} we introduce the necessary entropic quantities that come out in the expressions for finite key length secret key rates and derive the previously discussed non-asymptotic secret key rates. In Sec.~\ref{sec:OAM} we describe one possible laboratory implementation of our results by considering photonic OAM based QKD schemes, which have become a promising candidate for real-life high-dimensional QKD applications. We show the spatial modes that would be required for three MUBs and describe possible next steps and open challenges. We, however, do not analyze the security of the studied QKD protocols by considering all realistic parameters such a platform offers. This would include taking into account the efficiency of photon sources and detectors together with the suboptimality of certain classical information algorithms used in the postprocessing step. The experimental inefficiencies do not affect the secret key rate (measured by bits per channel) but rather the speed of how many secret bits one is able to collect per given time period.

\section{Security of asymptotic QKD and preliminaries}\label{sec:prelim}

The modern definition of security for a quantum key distribution protocol requires the final state $\vr_{ABE}$ to satisfy
\begin{equation}\label{eq:privacy}
  \big\|\vr_{ABE}-{1\over|K|}\sum_{k\in K}\kbr{k}{k}_A\otimes\kbr{k}{k}_B\otimes\tau_E\big\|_1\leq\e.
\end{equation}
The indices $A,B$ stand for the legitimate sender and receiver and $E$ is an adversary (Eve). The condition says that after the protocols ends, the legitimate parties share classical correlations (in this case a classical key $\{\kbr{k}{k}\}_k$), where the knowledge of Eve can be made arbitrarily small -- the quantum system in her possession is \emph{decoupled} from the legitimate participants. The expression $\|M\|_1\df\Tr{}\sqrt{MM^\dg}$ denotes the trace norm. This approach was first rigorously introduced in a great generality in~\cite{devetak2005private} and in the context of QKD also in~\cite{rennersecurity}. {The marginal state $\vr_{B}$ can be seen as an output of a noisy quantum channel $\Ncal$ between a sender and a receiver. They do not know whether the noisy evolution is caused by decoherence of any kind or by an eavesdropper and mainly they must not care. As long as they know the channel and are able to use it asymptotically (sending a large number of quantum signals) one can often easily determine whether a secret key can be established. Here comes the idea of asymptotic QKD: with an ever increasing number of channel uses the parameter on the RHS of Eq.~(\ref{eq:privacy}) is required to become arbitrarily small. For some channels this condition cannot ever be satisfied and in that case the asymptotic QKD is impossible. The normalized rate at which establishing classical correlation over a noisy quantum channel is in principle  possible   is called the \emph{private classical capacity} of $\Ncal$. Note that a secret key is a form of classical correlations~\cite{devetak2005private}. If the private capacity is zero, Eq.~(\ref{eq:privacy}) cannot be satisfied in the sense that Eve cannot be arbitrarily well decoupled from the state shared by the sender ($A$) to a receiver ($B$). The private classical capacity is given by}
\begin{equation}\label{eq:PrivCap}
  P(\Ncal)\df\lim_{n\to\infty}{1\over n}\sup_{\vr_{XA^n}}P(\Ncal^{\otimes n},\vr),
\end{equation}
where
\begin{equation}\label{eq:MutDifference}
  P(\Ncal,\vr)\df I(X;B)_\s-I(X;E)_\s
\end{equation}
is the \emph{private information}. The state $\s_{XBE}=\sum_xp_x\kbr{x}{x}\otimes\s_{x,BE}$ is given by the action of a channel isometry $W_\Ncal:A\mapsto BE$ on a classical-quantum input state $\vr_{XA}=\sum_xp_x\kbr{x}{x}\otimes\vr_{x,A}$ and $X$ denotes a classical random variable with a probability distribution $P$ ($p_x\equiv\Pr{(X=x)}$). The quantity $I(A;B)$ is called the \emph{quantum mutual information} defined as
\begin{equation}\label{eq:QuantMutInfo}
I(A;B)_\s=H(A)_\s+H(B)_\s-H(AB)_\s,
\end{equation}
where $H(A)_\s\df-\Tr{}[\s_A\log{\s_A}]$ is the von Neumann entropy\footnote{$\log{}$ is the logarithm base two and $\ln{}$ denotes the natural logarithm throughout the paper.} of a (possibly multipartite) state $\s_{AB\dots Z}$.  The private classical capacity in (\ref{eq:PrivCap}) is an unconstrained optimization problem whose tractable solution for a general channel $\Ncal$ is not known at present  and even the calculation of the \emph{one-shot} private capacity ($n=1$)
\begin{equation}\label{eq:oneshotPrivCap}
P^{(1)}(\Ncal)\df\sup_{\vr_{XA}}P(\Ncal,\vr)
\end{equation}
is not straightforward since $\vr_A$ admits a mixed state decomposition $\vr_{A}=\sum_xp_x\vr_{x,A}$.

Another fundamental quantity, seemingly unrelated to QKD, is called the \emph{quantum channel capacity}~\cite{devetak2005private}
\begin{equation}\label{eq:QuantCap}
  Q(\Ncal)\df\lim_{n\to\infty}{1\over n}\sup_{\vr_{A^n}}Q(\Ncal^{\otimes n},\vr),
\end{equation}
where
\begin{equation}\label{eq:CohInfo}
  Q(\Ncal,\vr)\df H(B)_\vt-H(E)_\vt
\end{equation}
is the \emph{coherent information}. The isometry now acts on $\vr_A$ that (crucially) can be limited to a convex sum of rank-one states $\om_{x,A}$ as $W_\Ncal:\sum_xp_x\kbr{\om_x}{\om_x}_{A}\mapsto\vt_{BE}$. The quantum capacity follows from a stronger condition than Eq.~(\ref{eq:privacy}) -- that the main goal is to successfully transmit a quantum state from a sender to a receiver who happens to be decoupled from the environment $E$ (completely controlled by an adversary). Quantum channel capacity~(\ref{eq:QuantCap}) is also intractable for a general channel $\Ncal$  but the one-shot quantity (also called the optimized coherent information)
\begin{equation}\label{eq:oneshotQuantCap}
  Q^{(1)}(\Ncal)\df\sup_{\vr_{A}}Q(\Ncal,\vr)
\end{equation}
is fairly easy to evaluate (often not analytically but the numerics will do the job).

{The decoupling mechanism is naturally useful for secret key generation. This is because the quantum capacity can crucially be interpreted as the one-way entanglement distillation rate which itself is a lower bound on the one-way secret key rate~\cite{devetak2005private}. Once the parties share maximally entangled states, they can be used to teleport any type of information, in particular a secret key, at the same rate the pairs were distilled. Hence the quantum capacity is a channel secret key rate lower bound. Formally, it can be shown in the following way}
~\cite{devetak2005private}  (see also~\cite{smith2008private}). From Eq.~(\ref{eq:oneshotPrivCap}) and the definition of the mutual information we write
\begin{subequations}\label{eq:PrivQuantCapRelation}
  \begin{align}%\label{}
   P^{(1)}(\Ncal) & = \sup_{\vr_{XA}}\big[I(X;B)_\s-I(X;E)_\s\big]\label{eq:PrivQuantCapRelationa}\\
     & = \sup_{\vr_{XA}}\big[H(B)_\s-H(BX)_\s-H(E)_\s+H(EX)_\s\big]\label{eq:PrivQuantCapRelationb}\\
     & = \sup_{\vr_{XA}}\big[H(B)_\s-H(E)_\s-\sum_xp_x\big(H(B)_{\s_{x}}-H(E)_{\s_{x}}\big)\big]\label{eq:PrivQuantCapRelationc}\\
     & = \sup_{\vr_A}\big[H(B)_\s-H(E)_\s\big]-\inf_{\vr_{XA}}\sum_xp_x\big(H(B)_{\s_{x}}-H(E)_{\s_{x}}\big)\label{eq:PrivQuantCapRelationd}\\
     & = Q^{(1)}(\Ncal)-\inf_{p_x,\vr_{x,A}}\sum_xp_xQ(\Ncal,\vr_x).)\label{eq:PrivQuantCapRelatione}
  \end{align}
\end{subequations}
Eq.~(\ref{eq:PrivQuantCapRelationc}) follows from
$$
H(BX)_\s=H\big(\sum_xp_x\kbr{x}{x}\otimes\s_{x,B}\big)=H(X)_P+\sum_xp_xH(B)_{\s_x}
$$
and similarly for the $H(EX)_\s$. The first two summands in Eq.~(\ref{eq:PrivQuantCapRelationd}) can be optimized over $\vr_A$ instead of $\vr_{XA}$ since we trace over the classical variable $X$. The von Neumann entropy $H(X)_P$ over a classical probability distribution $P$ is simply the Shannon entropy $S(\{p_x\})\df-\sum_xp_x\log{p_x}$. In the end we arrived at~\cite{devetak2005private}
\begin{equation}\label{eq:Oneshotinequality}
  Q^{(1)}(\Ncal)\leq P^{(1)}(\Ncal)
\end{equation}
and thus the LHS turns out to be a useful lower bound in the QKD scenario as claimed. The equality is achieved for $\vr_{x,A}=\kbr{\om_x}{\om_x}_{A}$ in which case $H(B)_{x,\vt}=H(E)_{x,\vt}$ for all $x$ and so \mbox{$Q(\Ncal,\vr_x)=0$}.

{The usual starting point for an asymptotic analysis of a QKD's secret key rate $r$ is the following formula~\cite{rennersecurity}\footnote{The actual expression for the key rate can be applied under very general circumstances, see~\cite{rennersecurity}, Corollary 6.5.2.}
\begin{equation}\label{eq:QKDsecretKeyRate}
  r_n\df{1\over n}\min_{\s_{AB}\in\Gamma}{\big[H(X^n|E^n)_\s-H(X^n|Y^n)_{\s}\big]},
\end{equation}
where $\s_{A^nB^nE^n}$ is a pure tripartite state shared by all parties, $\s_{XYE}$ is a classical-quantum state obtained by measuring $\s_{A^nB^nE^n}$ (so $X,Y$ are classical variables also called a raw key) and $n$ is the block size. The marginal state $\s_{A^nB^n}$ over which is being optimized is essentially a Choi state introduced on p.~\pageref{subsec:Jami}. The set $\Gamma$ are all Choi states compatible with the channel estimation step in the protocol and we will see it in action in Eqs.~(\ref{eq:CohInfoMinimizedMUB2c}).} Finally, the expression in Eq.~(\ref{eq:QKDsecretKeyRate})
\begin{equation}\label{eq:condEntrpy}
  H(A|B)_\vr\df H(AB)_\vr-H(B)_\vr
\end{equation}
is the quantum conditional entropy.  We can quickly see the equivalence between Eq.~(\ref{eq:QKDsecretKeyRate}) and $P^{(1)}(\Ncal)=\sup_{\vr_{XA}}[H(X|E)_\s-H(X|B)_\s]$ from  Eq.~(\ref{eq:PrivQuantCapRelationb}).  We  get rid of the supremum by realizing that in all mainstream QKD protocols, the input states (or private codes) $\vr_A$ are pure states (or mixtures thereof) leaving us with the classical-quantum input state of the form $\vr_{XA}=\sum_xp_x\kbr{x}{x}\otimes\kbr{\om_x}{\om_x}_{A}$. The maximum is achieved for  $\vr_{A}$ maximally mixed and so from Eq.~\ref{eq:PrivQuantCapRelatione} we get $P^{(1)}(\Ncal)=Q^{(1)}(\Ncal)$, see below~(\ref{eq:Oneshotinequality})\footnote{Note that we are not a priori assuming anything. If a new QKD protocol is invented, the fact that the one-shot private capacity is maximized for a maximally mixed state must be proved.}. In the second step, we realize that in all QKD protocols, Bob applies a POVM on the received quantum state generating a classical variable $Y$ and so Eq.~(\ref{eq:QKDsecretKeyRate}) for $n=1$ has been recovered
$$
Q^{(1)}(\Ncal)=r_1,
$$
where $\s_{AB}$ from the RHS represents $\Ncal$ on the LHS via the Choi-Jamio\l kowski isomorphism (see p.~\pageref{subsec:Jami}). There is also  a missing $\sup{}$ for $r_1$ (or $r_n$ in general) as opposed to $Q^{(1)}(\Ncal)$ and this a subtle point. From the quantum capacity standpoint, the channel $\Ncal$ is given and the maximization is over all possible input states $\vr_A$ (quantum codes). In the QKD scenario (specifically in its entanglement version) the parties try to share maximally entangled states and the most reasonable strategy is obviously to start the distribution with maximally entangled states (quantum codes)\footnote{A more general idea, that we will not discuss further, is the possibility  already envisaged in~\cite{devetak2005private} to go beyond entanglement distillation protocols in order to establish classical secret correlations. It indeed turns out that one can distribute so-called ``private states''~\cite{horodecki2005secure} for this purpose. This is precisely the situation where $Q^{(1)}(\Ncal)=0$ but $P^{(1)}(\Ncal)>0$.}. The fact that they may become disrupted due to decoherence or an eavesdropper implies that the channel will be different. As we will see later, such a disrupted code \emph{is} a channel representation (the Choi matrix).

\subsection*{Pauli group for qudits and MUBs}
It is instructive to investigate the case of two complementary bases (MUBs) for higher-dimensional Hilbert spaces. To this end, we first informally introduce the qudit Pauli group $\Pi_d$. It has two generators $X_d,Z_d\in\Pi_d$ defined as
\begin{subequations}\label{eq:PauliGens}
  \begin{align}%\label{}
    X_d & =\sum_{k=0}^{d-1}\kbr{k\oplus1}{k}, \\
    Z_d & =\sum_{k=0}^{d-1}\om^k\kbr{k}{k},
  \end{align}
\end{subequations}
where $\om=\exp{2\pi i/d}$ and $\oplus$ is addition modulo $d$. An arbitrary element of $\Pi_d$ is then $X_d^\a Z_d^\b$ for $0\leq\a,\b\leq d-1$.

From other useful properties of the qudit Pauli group let us recall that the special case of Weyl commutation relations~\cite{durt2010mutually}) reads
\begin{equation}\label{eq:PauliWeyl}
  X_dZ_d=Z_dX_d e^{i\zeta_d}.
\end{equation}
Hence,  the eigenvector $v_d$ in the equation  $X_dZ_dv_d=e^{i\la_d} v_d$  is also an eigenvector of $X_d^\a Z_d^\a$ (up to a phase).
This is because
\begin{equation}\label{eq:PauliCommutes}
    X_d^\a Z_d^\a=(X_dZ_d)^\a e^{i\ka\zeta_d},
\end{equation}
where $\ka=(\a^2-\a)/2$ counts the total number of passes of $Z_d$ through $X_d$. But $v_d$ is also an eigenvector of the RHS (up to a phase).

\subsection*{Choi-Jamio\l kowski representation of quantum channels}\label{subsec:Jami}

A remarkable way of representing a quantum channel is known as the Choi-Jamio\l\-kowski isomorphism~\cite{choi1975completely,jamiolkowski1972linear}. Let $\Ncal$ be the quantum channel. Then there exists a positive semi-definite map $R_\Ncal$, sometimes called \emph{Choi matrix},  that represents the action of the channel  via
\begin{equation}\label{eq:JamiChoi}
  \Ncal\circ\vr_A=\Tr{A}\big[(\vr_A^\top\otimes\id_{B})\,R_\Ncal\big].
\end{equation}
The channel $\Ncal$ is trace-preserving if its Choi matrix satisfies $\Tr{B}R_\Ncal=\id_A$. Conversely, any quantum channel $\Ncal$ gives rise to a Choi matrix
\begin{equation}\label{eq:ChoiConverse}
  R_{AB}(\Ncal)=(\id_A\otimes\,\Ncal)\circ\Phi_{AA'},
\end{equation}
where $\Phi_{AA'}=\sum_{i=1}^{d_A}\ket{i}_A\ket{i}_{A'}$ is an unnormalized maximally entangled state. The physical interpretation of the Choi matrix is as if the communicating parties shared a maximally entangled qudit pair. Instead of sending the actual qudit through the channel one sends a half of a maximally entangled state. The Choi matrix is usually derived from another channel representation (Kraus maps, for example) but almost all QKD schemes allow its direct construction. This leads to the so-called diagonal Bell state. To see this, recall that the states in many QKD schemes are always sent in one of the MUB bases. That means that the number of possible errors can be enumerated -- one just needs to find the error generators causing a bit flip in at least one of the bases. These are precisely the elements of the Pauli group $\Pi_d$ and so the Choi matrix reads
\begin{equation}\label{eq:QKDChoi}
  \tilde R_{AB}(\Ncal)=\sum_{\a,\b=0}^{d-1}\lambda_{\a\b}\big(\id\otimes X_d^\a Z_d^\b\big)\,\tilde\Phi_{AA'}.
\end{equation}
Starting from (\ref{eq:QKDChoi}), the operation $\circ$ in (\ref{eq:ChoiConverse}) becomes an ordinary matrix multiplication and the tilde indicates a normalized state. The probability error coefficients satisfy $1\geq\lambda_{\a\b}\geq0$ together with
$\sum_{\a,\b=0}^{d-1}\lambda_{\a\b}=1$.

\section{Derivation of the 2- and 3-MUB QKD adversarial channels for qudits and their asymptotic secret key rates}\label{sec:QKDasymptotic}

We adopt and reformulate the method of adversarial channel derivation from~\cite{rennersecurity}. A concise version also appears in Appendix~A of~\cite{scarani2009security}.

\subsection*{2 MUBs} 
\begin{figure}[h]
   \resizebox{10cm}{!}{\includegraphics{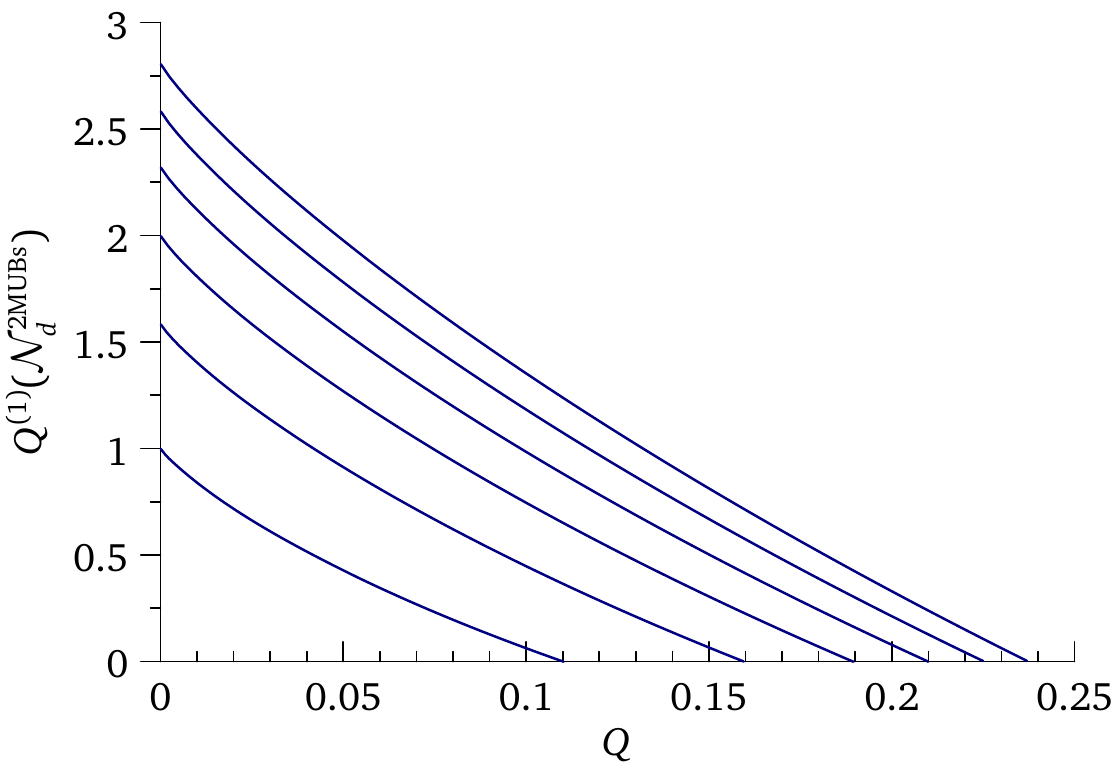}}
    \caption{Asymptotic secret key rates for 2-MUB QKD protocol (in bits per channel) are depicted for $d=2$ to $7$ (from the bottom up).}
    \label{fig:asympt2MUBs}
\end{figure}

The error analysis is straightforward. In the bit basis (the eigenvectors of $Z_d$), the errors are caused by $X^\a_d$ (there is $d-1$ of them) and $X^\a_dZ^\b_d$ for all $\a,\b>0$ (there is $(d-1)^2$ of them in total). Hence the measured error rate in the bit basis reads
\begin{equation}\label{eq:QBERbit2MUBS}
  Q_{b}=(d-1)\lambda_{Z}+(d-1)^2\lambda_{?},
\end{equation}
where $\la_Z\equiv\la_{0\b}$ and $\la_?$ is the rest. Similarly for the phase basis, by setting  $\la_X\equiv\la_{\a0}$ we obtain
\begin{equation}\label{eq:QBERphase2MUBS}
  Q_{p}=(d-1)\lambda_{X}+(d-1)^2\lambda_{?}.
\end{equation}
It is common and experimentally reasonable~\cite{scarani2009security} to set the error rates equal $Q_b=Q_p\equiv Q$. The normalization condition yields
\begin{equation}\label{eq:normalization2MUBS}
  {\la_{00}}=1-2Q+(d-1)^2\la_?
\end{equation}
and it is perhaps clear that $\la_?$ is a free parameter that needs to be determined by taking the best Eve's strategy. Following~\cite{rennersecurity}, the most general quantum attack is a collective attack. A collective attack is Eve's interaction with a passing qubit one by one with an eventual collective measurement deferred until the quantum transmission is over. In this light, the maximum amount of information provided to Eve is given by the \emph{minimized} coherent information Eq.~(\ref{eq:CohInfo}) which we readily rewrite as
\begin{equation}\label{eq:CohInfoI}
  Q(\Ncal,\tilde\Phi_{AA'})=H(B)_{\tilde R}-H(AB)_{\tilde R}.
\end{equation}
Indeed, the normalized Choi matrix $\tilde{R}$ serves a double purpose: it is a channel representation but also an output of the channel whose input is maximally entangled with the reference system $A$ (see Eq.~(\ref{eq:ChoiConverse})). The minimized RHS can be immediately evaluated
\begin{subequations}\label{eq:CohInfoMinimizedMUB2}
\begin{align}
  \min_{\la_?}Q(\Ncal,\tilde\Phi_{AA'})&=\min_{\la_?}{[H(B)_{\tilde R}-H(AB)_{\tilde R}]}\\
  &=\log{d}+\min_{\la_?}\sum_{\a,\b=0}^{d-1}\la_{\a\b}\log{\la_{\a\b}}\label{eq:CohInfoMinimizedMUB2b}\\
  &=\log{d}+\min_{\la_?}\big[
  (1-2Q+(d-1)^2\la_?)\log{[1-2Q+(d-1)^2\la_?]}\label{eq:CohInfoMinimizedMUB2c}\\
  &\ \quad\qquad+2(d-1){Q-(d-1)^2\la_?\over d-1}\log{Q-(d-1)^2\la_?\over d-1}\nn\\
  &\ \quad\qquad+(d-1)^2\la_?\log{\la_?}  \nn
  \big].
\end{align}
\end{subequations}
Equality (\ref{eq:CohInfoMinimizedMUB2b}) follows from $\Tr{A}[\tilde{R}_{AB}]=\id/d$ (the channel represented by $R_{AB}$ ($\tilde{R}_{AB}$) is unital). We also used the fact that $\tilde{R}_{AB}$ is Bell-diagonal in order to calculate $H(AB)_{\tilde R}$ using Eqs.~(\ref{eq:QBERbit2MUBS}),(\ref{eq:QBERphase2MUBS}) and (\ref{eq:normalization2MUBS}). {From~(\ref{eq:CohInfoMinimizedMUB2c}), by setting ${\dif{}\over\dif{\la_?}}[Q(\Ncal,\tilde\Phi_{AA'})]=0$, we find the stationary point
\begin{equation}\label{eq:optimalValue2MUBS}
  \la_?={Q^2\over(d-1)^2}
\end{equation}
and ${\diff{}\over\diff{\la_?}}[Q(\Ncal,\tilde\Phi_{AA'})]\big|_{(\ref{eq:optimalValue2MUBS})}=\frac{(d-1)^4}{(Q-1)^2 Q^2 \ln{2}}>0$ reveals a minimum for all $d$ and $Q$.} Then
\begin{equation}%\label{}
  \la_Z=\la_X={Q(1-Q)\over d-1}
\end{equation}
and as a result we get
\begin{align}\label{eq:quditOAMchannelGood}
  \Ncal_d^{\mathrm{2MUBs}}(\vr)&=(1-Q)^2\vr+{Q(1-Q)\over d-1}\sum_{\a=1}^{d-1}X_d^\a\vr {X_d^\a}^\dg+{Q(1-Q)\over d-1}\sum_{\b=1}^{d-1}Z_d^\b\vr {Z_d^\b}^\dg\nn\\
            &\quad+{Q^2\over (d-1)^2}\sum_{\a,\b=1}^{d-1}X_d^\a Z_d^\b\vr {\big(X_d^\a Z_d^\b\big)}^\dg
\end{align}
also called the BB84 channel for $d=2$. The secret key rates obtained by plugging Eq.~(\ref{eq:optimalValue2MUBS}) into Eq.~(\ref{eq:CohInfoMinimizedMUB2c}) read
\begin{equation}\label{eq:keyRatesBB84}
    Q^{(1)}(\Ncal_d^{\mathrm{2MUBs}})=\log{d}+2\big[Q\log{Q}+(1-Q)\log{(1-Q)}-Q\log{(d-1)}\big]
\end{equation}
and are plotted in Fig.~\ref{fig:asympt2MUBs} for $d=2$ to $7$. For $d=2$ the rate goes to zero for $Q\approx0.11$ which is the famous threshold derived in~\cite{shor2000simple}.

\subsection*{2 MUBs via Eq.~(\ref{eq:QKDsecretKeyRate})} We can recover one of our earlier results also from Eq.~(\ref{eq:QKDsecretKeyRate}). First, since $H(X)=H(B)=\log{d}$ and by using Eq.~(\ref{eq:QuantMutInfo}) together with the identity $H(B)-H(B|X)=H(X)-H(X|B)$  we get
\begin{equation}\label{eq:conditionalEntBB84B}
  H(X|B)=H(B|X)=-(1-Q)\log{(1-Q)}-Q\log{Q}+Q\log{(d-1)}.
\end{equation}
The channel  $\Ncal_d^{\mbox{\tiny 2MUBs}}$ is unital: $\Ncal_d^{\mbox{\tiny 2MUBs}}:{\id_d\over d}\mapsto{\id_d\over d}$. Therefore, Bob's information is classical (knowing the basis he perfectly measures the raw bit value), $Y\equiv B$ and $H(X|B)=H(X|Y)$. We also find
$$
    {H(E)\over2}=H(E|X)\equiv H(B|X)
$$
and by using $H(E)-H(E|X)=H(X)-H(X|E)$ we get
\begin{equation}\label{eq:conditionalEntBB84E}
  H(X|E)=\log{d}-H(E|X).
\end{equation}
Putting it all together, we obtain
\begin{equation}\label{eq:QKDsecretKeyRateBB84Again}
  {r_1^{(d,\mathrm{2MUBs})}}=\log{d}+2\big[(1-Q)\log{(1-Q)}+Q\log{Q}-Q\log{(d-1)}\big]\equiv Q^{(1)}(\Ncal_d^{\mathrm{2MUBs}})
\end{equation}
in accordance with Eq.~(\ref{eq:keyRatesBB84}).

The reason for the repetition of the previous analysis is two-fold. Besides showing that our earlier approach via quantum/private capacity is valid and arguably more perspicuous, the secret key rates of the form of Eq.~(\ref{eq:QKDsecretKeyRate}) enable a nice interpretation of the entropic quantities and a direct comparison with the results coming from the finite key size analysis performed in~\cite{tomamichel2012tight}, which is based on the one-shot entropic uncertainty relations. The second point will be discussed in detail in Section~\ref{sec:finiteSizeKey}. To illustrate the first point, note that for $d=2$ we may rewrite Eq.~(\ref{eq:QKDsecretKeyRate}) in an even more familiar form~\cite{scarani2009security}
\begin{equation}\label{eq:QKDsecretKeyRateBB84Simple}
    {r_1^{(d,\mathrm{2MUBs})}}=1-h(Q)-\mathrm{leak_{EC}},
\end{equation}
where $h(Q)\df-(1-Q)\log{(1-Q)}-Q\log{Q}$ is the binary Shannon entropy and $\mathrm{leak}_{EC}=h(Q)$ is the information leaked to Eve during the error correction (information reconciliation) procedure.

{Going back to a general $d$, typically,  $\mathrm{leak}_{EC}>H(X|Y)$ (recall $Y\equiv B$ from below Eq.~(\ref{eq:conditionalEntBB84B})). This is because the algorithms performing this purely classical part do  not typically achieve the Shannon limit~\cite{scarani2008quantum}. For our purposes we consider this step to be perfect: $\mathrm{leak}_{EC}=H(X|Y)$.}

\subsection*{3 MUBs} The existence of three MUBs  generated by the Pauli elements $Z_d ,X_d$ and $X_dZ_d$  for any $d$~\cite{durt2010mutually} is good news and it makes senses to study the secret key rates for the 3-MUB QKD protocols. The error analysis is a bit more intricate. In the bit ($Z_d$) and phase ($X_d$) basis the errors are generated by the $X_d^\a$ and $X_d^\a Z_d^\b$ and by $Z_d^\b$ and $X_d^\a Z_d^\b$, respectively, assuming $\a,\b>0$. In the bit-phase basis (the basis spanned by the eigenvectors of $X_dZ_d$) the errors are caused by $X_d^\a,Z_d^\b$ ($\a,\b>0$) and those \emph{not} of the form $X_d^\a Z_d^\a$ for $\a>0$. This is shown in Eq.~(\ref{eq:PauliCommutes}).

\begin{figure}[h]
   \resizebox{10cm}{!}{\includegraphics{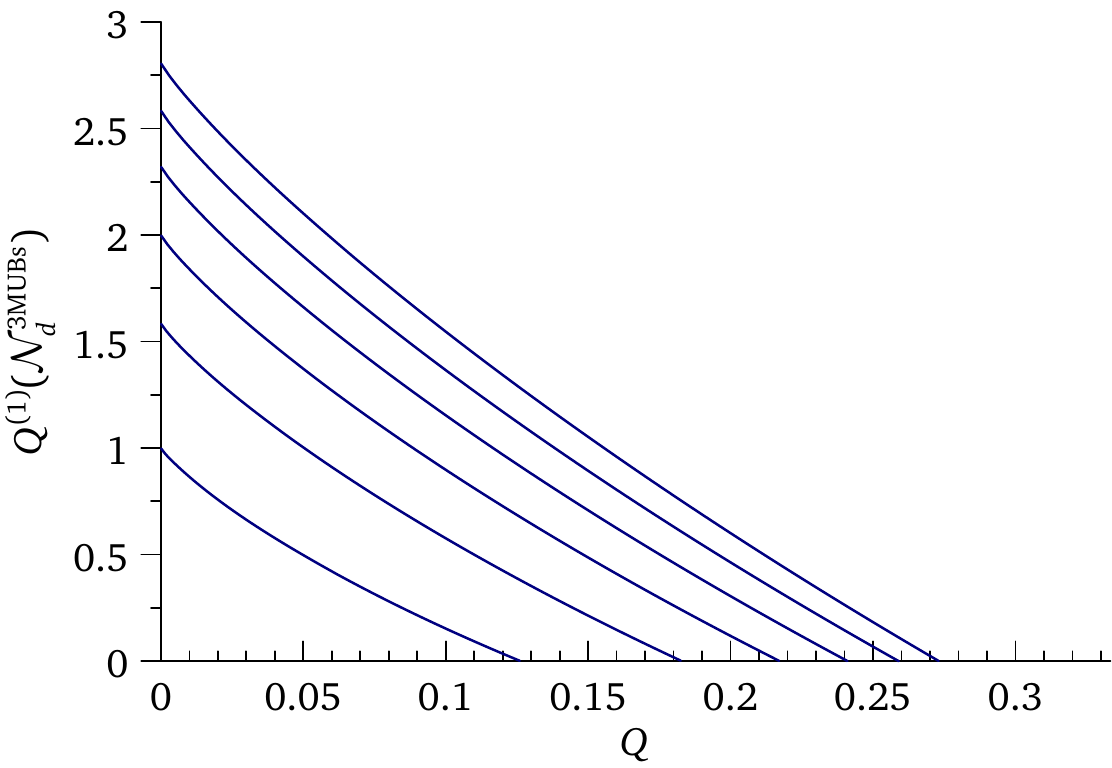}}
    \caption{Qudit  secret key rates  for the 3-MUB QKD protocol for $d=2$ (the bottom curve) to $7$ (the upmost curve) in bits per channel are plotted. For the special case $d=2$  Eqs.~(\ref{eq:QBER_3MUBS}) simplify and no optimization is needed. The resulting channel is Eq.~(\ref{eq:6stchannel})}
    \label{fig:asympt3MUBs}
\end{figure}

Let us first do some counting: for a given $d$ there is in total $d^2-1$ error operators $X_d^\a Z^\b_d$ by excluding an identity. It contains $d-1$ of $X_d^\a$ operators and $d-1$ of $Z_d^\b$ operators. There is also $d-1$ of $X_d^\a Z_d^\a$ operators for $\a>0$. Hence, the number of operators of the form $X_d^\a Z_d^\b$ ($\a,\b>0,\a\neq\b$) causing errors in the $X_dZ_d$ basis must be
$$
d^2-1-3(d-1)=(d-2)(d-1).
$$
As a result we get from Eq.~(\ref{eq:QKDChoi}) the following error rates:
\begin{subequations}\label{eq:QBER_3MUBS}
\begin{align}
  Q_{b}&=(d-1)\la_{Z}+(d-1)\la_X+(d-2)(d-1)\la_{XZ},\\
  Q_{p}&=(d-1)\la_{Z}+(d-1)\la_?+(d-2)(d-1)\la_{XZ},\\
  Q_{b-p}&=(d-1)\la_{X}+(d-1)\la_?+(d-2)(d-1)\la_{XZ}.
\end{align}
\end{subequations}
The coefficients $\la_Z,\la_X$ are defined as before and $\la_{XZ}=\la_{\a\a}$ for $0<\a\leq d-1$. We again set the error rates equal: $Q_b=Q_p=Q_{b-p}\equiv Q$. The normalization condition becomes
\begin{equation}%\label{}
  {\la_{00}}+(d-1)\la_Z+(d-1)\la_X+(d-1)\la_?+(d-2)(d-1)\la_{XZ}=1
\end{equation}
and we find
\begin{subequations}\label{eq:lambdaCoeffsMUB3}
  \begin{align}%\label{}
    {\la_{00}} &= 1-Q-(d-1)\la_?, \\
    \la_X &= \la_Z = \la_?,\\
    \la_{XZ} &= {Q-2(d-1)\la_?\over(d-2)(d-1)}
  \end{align}
\end{subequations}
for $d>2$. The channel is of the following form
\begin{align}\label{eq:qudit3MUBS}
  \Ncal_d^{\mathrm{3MUBs}}(\vr)&=(1-Q-(d-1)\la_?)\vr+\la_?\Big[\sum_{\a=1}^{d-1}X_d^\a\vr {X_d^\a}^\dg+\sum_{\b=1}^{d-1}Z_d^\b\vr {Z_d^\b}^\dg+\sum_{\g=1}^{d-1}X_d^\g Z_d^\g\vr {\big(X_d^\g Z_d^\g\big)}^\dg\Big]\nn\\
            &\quad+{Q-2(d-1)\la_?\over(d-2)(d-1)}\sum_{\a\neq\b=1}^{d-1}X_d^\a Z_d^\b\vr {\big(X_d^\a Z_d^\b\big)}^\dg.
\end{align}
The minimization procedure similar to Eq.~(\ref{eq:CohInfoMinimizedMUB2}) leads to an analytical solution (too long to paste here) of the following cubic equation
\begin{equation}\label{eq:optimalValue3MUBS}
  \la^3_?=\big(1-(d-1)\la_?-Q\big)\bigg[{-2(d-1)\la_?+Q\over(d-2)(d-1)}\bigg]^2.
\end{equation}
The resulting secret key rates are given by
\begin{align}\label{eq:keyRates3MUBS}
    Q^{(1)}(\Ncal_d^{\mathrm{3MUBs}}) &=\log{d}+(Q-2(d-1)\la_?)\log{Q-2(d-1)\la_?\over(d-2)(d-1)}\nn\\
    &\quad +3(d-1)\la_?\log{\la_?}+(1-Q-(d-1)\la_?)\log{(1-Q-(d-1)\la_?)}
\end{align}
and are plotted in Fig.~\ref{fig:asympt3MUBs}. By comparing with Fig.~\ref{fig:asympt2MUBs} we can see that the tolerable threshold values are much better than for the corresponding 2-MUB protocol. {Our results perfectly agree (in  the overlapping cases) with a numerical study from~\cite{coles2015unstructured} as well as the secret key rates and thresholds from~\cite{ferenczi2012symmetries}.}
The $d=2$ case must be analyzed separately and it is the well-known six-state protocol~\cite{bruss1998optimal}. The channel is the qubit depolarizing channel (see again~\cite{scarani2009security}, Appendix~A)
\begin{equation}\label{eq:6stchannel}
  \Ncal_2^{\mathrm{3MUBs}}(\vr)=(1-3Q/2)\vr+Q/2(X_2\vr X_2+Y_2\vr Y_2+Z_2\vr Z_2).
\end{equation}
Then
\begin{equation}\label{eq:BB84errorrate}
  Q^{(1)}(\Ncal_2^{\mathrm{3MUBs}})=1-S(\{q_i\}),
\end{equation}
where $q_i=\{1-3/2Q,Q/2,Q/2,Q/2\}$. The one-shot capacity becomes zero for the threshold value $Q\approx0.126$~\cite{lo2001proof,scarani2009security}.

\section{Non-asymptotic secret key rates for the 2- and 3-MUB QKD $d$-level protocols}\label{sec:finiteSizeKey}

{The condition for a secret key generated when the resources  are not unlimited is formally identical to Eq.~(\ref{eq:privacy}). However, Eq.~(\ref{eq:privacy}) cannot this time be satisfied arbitrarily well. More precisely, for finite-length private codes, $\e$ is chosen sufficiently small and it becomes an input parameter of the secret key generation protocol. The task can be further reformulated --  it is often advantageous to investigate separately two conditions : (i) \emph{correctness}
\begin{equation}\label{eq:correctness}
  \mathrm{Pr}[K_A\neq K_B]\leq\e_{cor},
\end{equation}
where the key string is allowed to be different with a nonzero probability $\e_{cor}$, and (ii) \emph{secrecy}
\begin{equation}\label{eq:secrecy}
  \big\|\vr_{AE}-{1\over|K|}\sum_{k\in K}\kbr{k}{k}_A\otimes\tau_E\big\|_1\leq\e_{sec}.
\end{equation}
This means that an adversary is decoupled from the resulting secret key sequence by a small (but fixed) amount $\e_{sec}$.} Due to composability~\cite{muller2009composability}, the errors add up and the overall security parameter is bounded: $\e\leq\e_{cor}+\e_{sec}+\e_{PA}$\footnote{\label{foot:relQKD}We took the liberty of ignoring the possibility of failure $\e_{PA}$ during the privacy amplification (PA) step and the probability of failure $\e_{cor}$ of correctly estimating Alice's key, Eq.~(\ref{eq:correctness}). Both parameters are undoubtedly important for the overall secret key rate in the non-asymptotic scenario. They manifest themselves as additional exponents in Eq.~(\ref{eq:leftover}) in the form proportional to $-\log{[1/\e]}$. The errors are chosen independently  as part of the protocol~\cite{scarani2008quantum,tomamichel2012tight} but our main interest lies in $\e_{sec}$ and so we will study the key rate as its function. For a practical piece of advice as what to do in the deployed scenario, where all parameters must be set, we point the reader to~Ref.~\cite{cai2009finite} and also~\cite{tomamichel2015rigorous}.}. Similarly to the asymptotic analysis, the ``measure'' of decoupling, $\e_{sec}$, is related, through the decoupling lemma~\cite{rennersecurity}
\begin{equation}\label{eq:leftover}
  \e_{sec}\leq2\ve+2^{-{1\over2}\left(-\ell+\hmineps(X^n|E^n)_\vr-n\mathrm{leak}_\mathrm{EC}\right)},
\end{equation}
to  the smooth min-entropy
\begin{equation}\label{eq:smoothminepsEnt}
  \hmineps(A|B)_\vr\df\max_{\genfrac{}{}{0pt}{2}{\vr'\mathrm{\ s.t.}}{\|\vr-\vr'\|_1\leq\ve}}\hmin(A|B)_{\vr'},
\end{equation}
 where
\begin{equation}\label{eq:minEnt}
  \hmin(A|B)_{\vr}\df\max_{\genfrac{}{}{0pt}{2}{\s_B\mathrm{\ s.t.}}{0<\Tr{}\s_B\leq1}}{\sup_{\xi\in\bbR}{[\vr_{AB}-2^{-\xi}\id_A\otimes\s_B\leq0]}}.
\end{equation}
We will also need the max-entropy definition
\begin{equation}\label{eq:maxEnt}
  \hmax(A|B)_\vr\df\sup_{\s_B}\log{\Big[\Tr{}\big[\big(\sqrt{\vr_{AB}}(\id_A\otimes\s_B)\sqrt{\vr_{AB}}\big)^{1/2}\big]\Big]^2},
\end{equation}
where for two commuting distributions $\vr\to P$ and $\s\to Q$ the optimization can be performed~\cite{tomamichel2012framework}.

Given the secrecy parameter $\e_{sec}$, the secret key of the length $\ell\df n r^{(\ve,n)}$ can be extracted whenever
\begin{equation}\label{eq:secrecyCond}
  r^{(\ve,n)}\leq{1\over n}\hmineps(X^n|E^n)_\vr-\mathrm{leak_{EC}}.
\end{equation}
{The secret key rate is achievable~\cite{privComRenner}. Given the security parameters $\ve$ in~(\ref{eq:leftover}), the constructed code satisfies the decoupling condition. In coding theory, the statement of achievability is usually proved by a random construction via a direct coding theorem. This is precisely the construction found in Sec.~5.4~of~\cite{rennersecurity}.} The original derivation from~\cite{rennersecurity} has been further elaborated on and sharpened providing increasingly better estimates for the secret key rate. For the most important contributions, we should not forget to mention~\cite{renner2005information,scarani2008quantum,abruzzo2011quantum,bratzik2011min} and mainly~\cite{tomamichel2012tight} culminating in~\cite{tomamichel2015rigorous} whose extension to the QKD qudit protocols will be presented in the next section. Also note the similarity between Eq.~(\ref{eq:QKDsecretKeyRate}) and Eq.~(\ref{eq:secrecyCond}). Indeed, this is not a coincidence, the latter can be seen as a finite-key version of the former~\cite{rennersecurity,scarani2008quantum}.  The conditional entropy belongs to a parametric family of the so-called R\'enyi entropies and both the min- and max-entropy, Eq.~(\ref{eq:minEnt}) and~\ref{eq:maxEnt}, are family members with an operational meaning relevant for QKD~\cite{renner2005information}. Furthermore, we have the equipartition property
\begin{equation}\label{eq:equipartition}
  \lim_{\ve\to0}\lim_{n\to\infty}{1\over n}\hmaxeps(X^n|E^n)_\vr=\lim_{\ve\to0}\lim_{n\to\infty}{1\over n}\hmineps(X^n|E^n)_\vr=H(X|E)_\vr.
\end{equation}

An important advance in the security of the finite-key size QKD using two MUBs was possible due to the use of the uncertainty relation for smooth entropies:
\begin{equation}\label{eq:uncertRel}
  \hmineps(X^n|E^n)_\vr+\hmaxeps(X^n|Y^n)_\vr\geq n\log{d}
\end{equation}
Ref.~\cite{tomamichel2012framework} explains the physical interpretation in detail so we will only say that uncertainty relations in general limit the knowledge in one basis if a measurement is performed in the complementary basis. In this case, the complementary basis (the eigenvectors of the Pauli $Z_d$ basis) is used exclusively for the sacrificed portion of the sifted key and this consequently serves for an estimation of the preserved part of the sifted key (which itself is transmitted in the basis spanned by the eigenvectors of the Pauli $X_d$ matrix).

\subsection*{Bounds on the finite secret key rate}

The direct evaluation of the smooth min-entropy for $0\ll n<\infty$ in Eq.~(\ref{eq:secrecyCond}) is not straightforward. There exists a couple of methods to estimate it and the most advanced analysis so far, based on the smooth entropy uncertainty relations, appeared in~\cite{tomamichel2015rigorous} following~\cite{tomamichel2012tight}. We present its generalization to the qudit scenario for the 2-MUB QKD protocol. This approach provides the best secret key rate known to the authors but it cannot be extended to the case of 3 MUBs in a straightforward manner. In this case we use another strategy via the study of the asymptotic behavior of the smooth min-entropy. This bound already appeared in~\cite{renner2005information} and we improve it by recent insights based on the \emph{conditional entropy variance} (the so-called second-order approximation of the quantum coding rate~\cite{tomamichel2015quantum}). For the sake of comparison, we evaluate these bounds also for the 2-MUB qudit protocol. Here, the finite-key corrections come from two sources. First, it is the approximations of the smooth min-entropy and the smooth quantities in general. The second source of corrections is  the error rate estimation phase, where a part of the sifted key is sacrificed in order to estimate the error rate of the data used to extract the actual secret key.

To proceed, we will recapitulate the relevant parts\textsuperscript{\ref{foot:relQKD}} of the qudit 2- and 3-MUB QKD protocol in order to apply the methods of~\cite{tomamichel2012tight}. For the case of 2 MUBs, we may adopt the same protocol definition as in Box~1 of~Ref.~\cite{tomamichel2012tight}. {In particular, an asymmetric choice of the complementary bases is used~\cite{lo2005efficient}, one for the raw key whose length  will be labeled $n$ and the other one of the length $k$ used solely in the parameter estimation step. Hence, the total length of the sifted key is $N=n+k$.} The  difference compared to~\cite{tomamichel2012tight}  is the calculation of the average error $\la$ subsequently used for the parameter estimation. As a pure formality -- instead of the modulo two addition of the publicly announced bit sequences of the length~$k$ (used to count the number of differing bits), the communicating parties may use
\begin{equation}\label{eq:errorCount}
  \la\df\sum_{i=1}^k\mathsf{\mathbf{1}}\{x_i\neq y_i|X\}=\sum_{i=1}^k\bigg\lceil{x_i\ominus y_i\over d}\bigg\rceil,
\end{equation}
where $\ominus$ stands for the modulo $d$ subtraction and $\mathsf{\mathbf{1}}\{\om|A\}$ denotes the set indicator function defined for two sets $\Omega\subset A$ as $\mathsf{\mathbf{1}}\{\om|A\}=1$ whenever $\om\in\Omega$ and zero otherwise. In the parameter estimation phase, the sacrificed portion of the sifted sequence of the length $k$  over $d$ letters (transmitted in the Pauli $Z_d$ basis) is used to estimate the error rate in the portion of the length $n$ transmitted in the Pauli $X_d$ basis. Analogously to~\cite{tomamichel2012tight}, we are penalized by effectively increasing the error rate  by $\nu=\sqrt{\frac{N(k+1)\ln{\frac{2}{\ve}}}{k^2 (N-k)}}$ due to the finiteness of the statistics. More precisely, the estimate of large deviations for an independent and identically distributed random process sampled without replacement due to Serfling is used~\cite{serfling1974probability}.

For three MUBs, the QKD protocol must be modified only such that the Pauli $X_d$ basis will be used for the key extraction and the $Z_d$ and $X_dZ_d$ basis for the parameter  estimation step. So the communicating parties will be instructed to switch the bases accordingly with equal probabilities for the $Z_d$ and $X_dZ_d$ bases. In this case, the uncertainty relations based approach does not provide the best secret key rates and the smooth min-entropy from Eq.~(\ref{eq:leftover}) must be estimated differently (see Eq.~(\ref{eq:smoothMinEntEstimateRenner}) onwards).

A useful {upper bound} on the classical max-entropy is given by the probability distribution support (the set over which the probability distribution is positive~\cite{rennersecurity}) leading to
\begin{equation}\label{eq:maxEntclassicalBound}
  \hmax(X|Y)_P\leq\log{|\supp{[P(X|Y=y)]}|}=\log{\big|\big\{x\in\{0,1,\dots,d-1\};\Pr{[X=x|Y=y]}>0\big\}\big|}.
\end{equation}
Here we generalize the result from~\cite{tomamichel2015rigorous} (Claim~9) and show that the RHS satisfies
\begin{equation}\label{eq:supportProbBound}
  \log{\big|\big\{x\in\{0,1,\dots,d-1\};\Pr{[X=x|Y=y]}>0\big\}\big|}\leq n\big(h(Q+\nu)+(Q+\nu)\log{(d-1)}\big)
\end{equation}
for the 2-MUB protocol. We start as in~\cite{tomamichel2012tight}
\begin{subequations}\label{eq:HminEstimate}
\begin{align}
  \big|\big\{x\in\{0,1,\dots,d-1\};\Pr{[X=x|Y=y]}>0\big\}\big|
        & \leq\sum_{x\in\{0,\dots,d-1\}^n}\mathsf{\mathbf{1}}\{\la<n(Q+\nu)\} \\
   & = \sum_{\la=0}^n\binom{n}{\la}(d-1)^\la\mathsf{\mathbf{1}}\{\la<n(Q+\nu)\}\\
   & = \sum_{\la=0}^{n(Q+\nu)}\binom{n}{\la}(d-1)^\la\\
   & \leq 2^{n(h(Q+\nu))}(d-1)^{n(Q+\nu)}.
\end{align}
\end{subequations}
The new term $(d-1)^\la$ in the first equality comes from an additional number of errors caused by a larger ($d$-letter) alphabet. The last line comes from  $\sum_{\la=0}^{n(Q+\nu)}\binom{n}{\la}\leq2^{n(h(Q+\nu))}$, valid for $0\leq Q+\nu\leq1/2$, and by taking into account $0\leq\la\leq n(Q+\nu)$. Upon taking the logarithm we obtain~(\ref{eq:supportProbBound}). This, on the other hand, allows us to bound the min-entropy from Eq.~(\ref{eq:leftover}) via~Eq.~(\ref{eq:uncertRel}):
\begin{equation}\label{eq:smoothMinEntEstimateUncert}
  \hmineps(X^n|E^n)_\vr\geq n\big(\log{d}-h(Q+\nu)-(Q+\nu)\log{(d-1)}\big).
\end{equation}
Hence, we get for~(\ref{eq:secrecyCond})
\begin{equation}\label{eq:secrecyCondI}
  r^{(\ve,n)}\leq\log{d}-h(Q+\nu)-(Q+\nu)\log{(d-1)}-\mathrm{leak_{EC}}
\end{equation}
and so finally the optimized secret key rate is given by
\begin{equation}\label{eq:KeyRateFiniteI}
  \wh{r}^{(\ve,n)}\leq \max_k{N-k\over N}\big[ \log{d}-h(Q+\nu)-(Q+\nu)\log{(d-1)}-\mathrm{leak_{EC}} \big].
\end{equation}
{The numerical optimization was done by choosing a target number of sifted signals $N$, the error rate $Q$ and the security parameter $\ve$. The result of optimization is the highest rate and also the number $k$ of sacrificed bits needed to achieve it. Another option, we did not pursue, was to set the target number $n$ of raw bits and optimize the rate over $k$ sa well. The choice depends more on practical requirements. As expected, in the limit of $N\to\infty$ or $n\to\infty$, we recover Eq.~(\ref{eq:keyRatesBB84}) ((\ref{eq:QKDsecretKeyRateBB84Again})). This is because $\nu\to0$ and ${N-k\over N}={n\over n+k}\to1$.}

The smooth min-entropy estimates reveal the rate of convergence in Eq.~(\ref{eq:equipartition}). The first such estimate widely used in the literature was provided by Renner~\cite{rennersecurity} (Cor.~3.3.7)
\begin{equation}\label{eq:smoothMinEntEstimateRenner}
  {1\over n}\hmineps(X^n|E^n)_\vr\geq H(X|E)_\vr-(2\log{\rank{\vr_X}}+3)\sqrt{{1\over n}\log{2\over\ve}}.
\end{equation}
A better estimate comes from the recent advances in finite block length quantum coding~\cite{tomamichel2015quantum} through
\begin{equation}\label{eq:smoothMinEntEstimateToma}
{1\over n}\hmineps(X^n|E^n)_\vr\geq H(X|E)_\vr+\Phi^{-1}(\ve^2)\sqrt{V(X|E)\over n},%+\Ocal(\log{(n)}/n),
\end{equation}
where
\begin{equation}\label{eq:RelVariance}
  V(\vr\|\s)\df\Tr{}\big[\vr\big(\log{\vr}-\log{\s}-D(\vr\|\s)\big)^2\big]
\end{equation}
is the relative entropy variance and
\begin{equation}\label{eq:RelEntropy}
  D(\vr\|\s)\df\Tr{}\big[\vr\big(\log{\vr}-\log{\s}\big)\big]
\end{equation}
is the quantum relative entropy~\cite{umegaki1962conditional}. Then, as a special case, we  obtain the quantum conditional entropy and the conditional entropy variance~\cite{li2014second}
\begin{align}\label{eq:conditionalStuff}
   H(A|B)_\vr &= -D(\vr_{AB}\|\id_A\otimes\vr_B) ,\\
   V(A|B)_\vr &= V(\vr_{AB}\|\id_A\otimes\vr_B).
\end{align}
The expression $\Phi^{-1}(x)=-\sqrt{2}\ \mathrm{inv}\,[(1-\mathrm{Erf}\,(2x))]$ stands for the inverse of the complementary cumulative Gaussian distribution function. The previously mentioned large deviation estimate of the smooth min-entropy manifests itself by replacing $f(Q)=H(X|E)_\vr$ with
\begin{equation}\label{eq:fluct}
  f(Q+\nu)=\widetilde{H}(X|E)_\vr\leq f(Q)
\end{equation}
in Eqs.~(\ref{eq:smoothMinEntEstimateRenner}) and (\ref{eq:smoothMinEntEstimateToma}).
\begin{figure}[h]
   \resizebox{10cm}{!}{\includegraphics{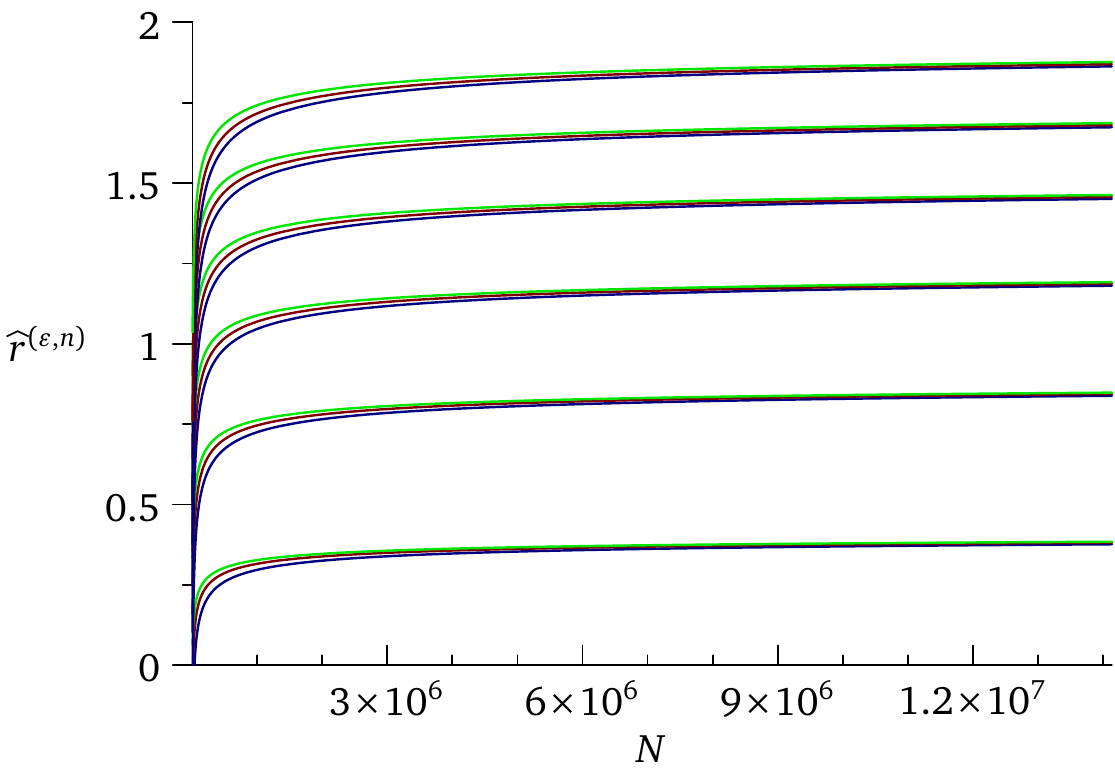}}
    {\caption{Secret key rates based on the finite-key length analysis for $d=2\dots7$ 2-MUB QKD protocol. For each $d$,  a triple of curves (blue/red/green) corresponds to increasingly better key rates. The worst rate (blue) is provided  by optimizing the lower expression in Eq.~(\ref{eq:KeyRateFinite}). The middle (red) curve comes from the second-order  analysis in the upper expression~Eq.~(\ref{eq:KeyRateFinite}). The highest (green) rate is given by optimizing Eq.~(\ref{eq:KeyRateFiniteI}) based on uncertainty relation for smooth entropies we obtained for any~$d$. We set $Q=0.05$, $\ve=10^{-10}$ and $N=n+k$ is the length of the sifted string of $d$ letters.}\label{fig:finitekey2MUBS} }
\end{figure}
\begin{figure}[h]
   \resizebox{10cm}{!}{\includegraphics{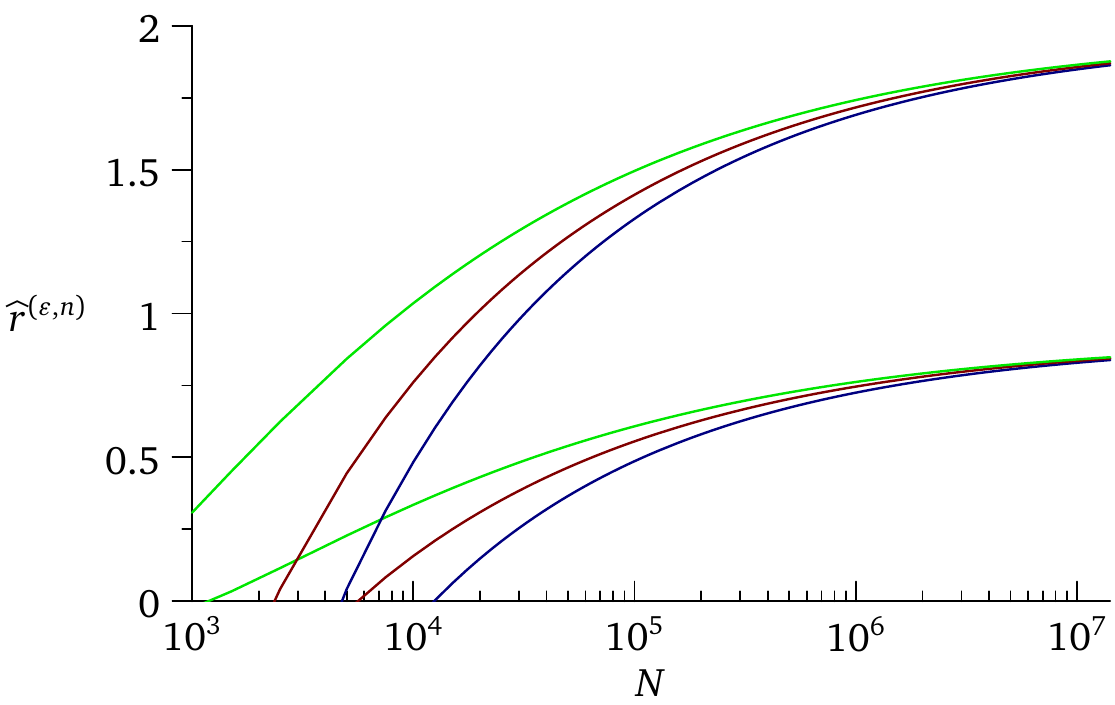}}
    \caption{Rescaled secret key rates from Fig.~\ref{fig:finitekey2MUBS} for $d=3$ and $d=7$ (using the same color coding) to assess the behavior for a low number of signals and show the superior rates provided by the uncertainty-relations-based approach (the green curves).}
    \label{fig:finitekey2MUBSlog}
\end{figure}
\begin{figure}[h]
   \resizebox{10cm}{!}{\includegraphics{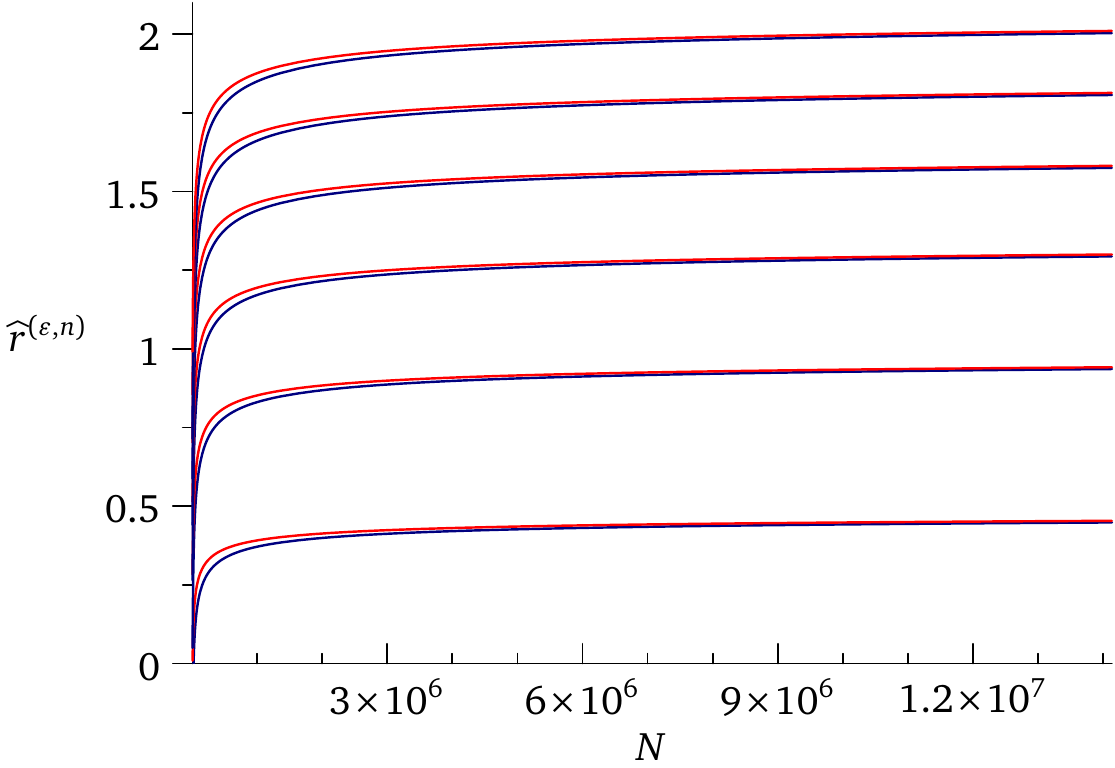}}
    {\caption{Secret key rates based on the finite-key length analysis for 3-MUB QKD protocol for $d=2\dots7$. The upper curve of each pair (red/blue) is given by optimizing the lower expression in Eq.~(\ref{eq:KeyRateFinite}). Hence the second-order  analysis provides better achievable rates compared to Renner's original estimate~\cite{rennersecurity} (lower curves from the upper expression in Eq.~(\ref{eq:KeyRateFinite})). We set $Q=0.05$, $\ve=10^{-10}$  and $N=n+k$ is the length of the sifted string of $d$ letters.  }  \label{fig:finitekey3MUBS} }
\end{figure}

Combining Eq.~(\ref{eq:secrecyCond}) and the estimates in Eqs.~(\ref{eq:smoothMinEntEstimateRenner}) and~(\ref{eq:smoothMinEntEstimateToma}) together with Eq.~(\ref{eq:fluct}) we get an achievable upper bound for the secret key rate
\begin{equation}\label{eq:KeyRateFinite}
  \wh{r}^{(\ve,n)}\leq \max_k{N-k\over N}\Bigg[\widetilde{H}(X|E)-\mathrm{leak}_\mathrm{EC}-
  \begin{cases}
    & (2\log{\rank{\vr_X}}+3)\sqrt{{1\over N-k}\log{2\over\ve}}\\
    & -\,\Phi^{-1}(\ve^2)\sqrt{V(X|E)\over N-k}.
  \end{cases}
  \Bigg].
\end{equation}
The optimized secret key rate $\wh{r}^{(\ve,n)}$ is plotted as the two lower curves in Fig.~\ref{fig:finitekey2MUBS} for the 2-MUB protocol and in Fig.~\ref{fig:finitekey3MUBS} for the 3-MUB protocol. Then, the overall number of secret key bits is given by $(N-k)\log{d}$ for $k$ found in Eq.~(\ref{eq:KeyRateFinite}). Fig.~\ref{fig:finitekey2MUBSlog} shows the $d=3$ and $d=7$ cases from Fig.~\ref{fig:finitekey2MUBS} on a semilogarithmic scale.

\section{Discussion and Conclusions}\label{sec:OAM}
\begin{figure}[t]
   \centering{\includegraphics[width=11cm]{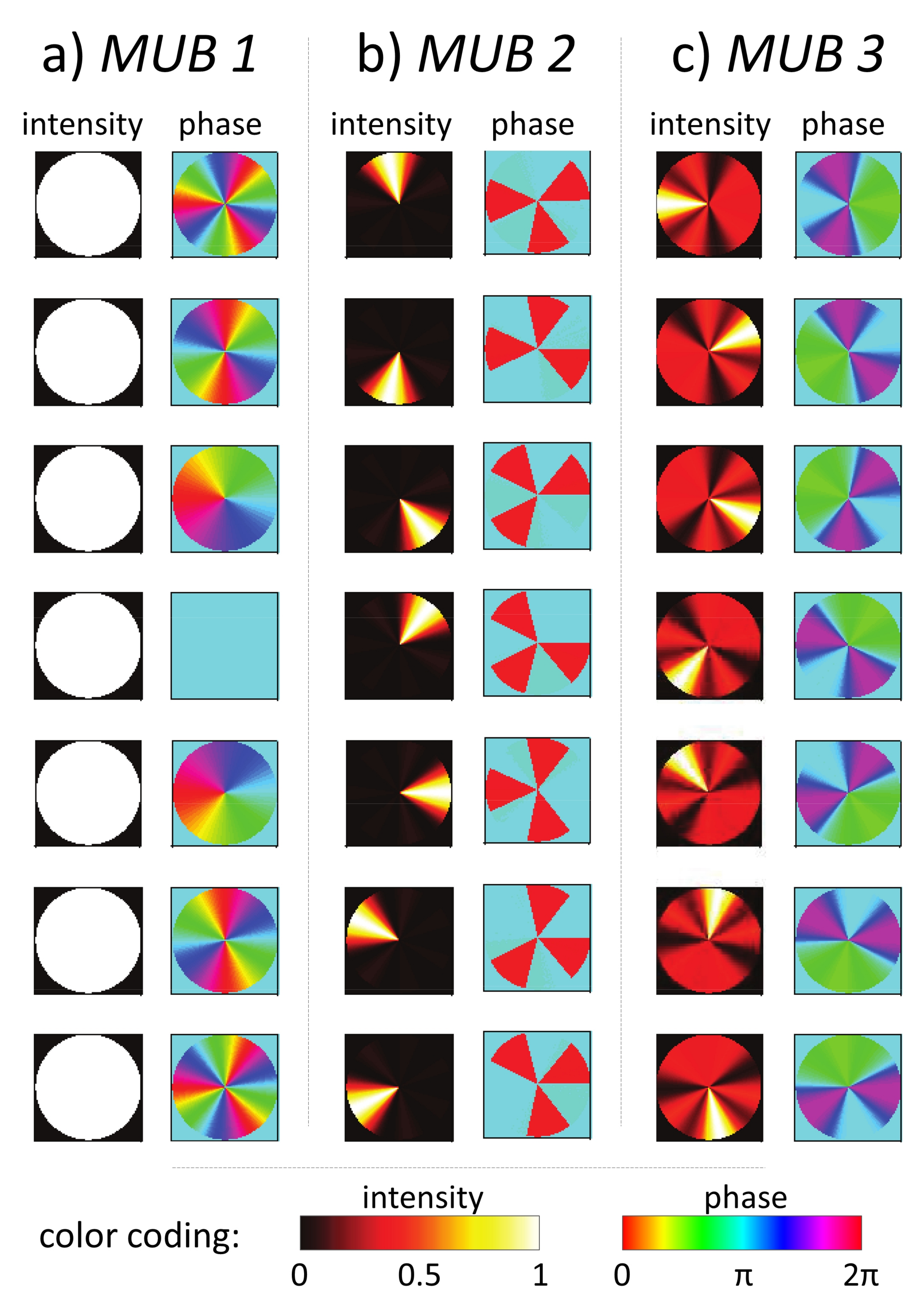} }
    \caption{{Normalized intensity (left columns) and  the corresponding phase plots (right columns) of the three 	mutually unbiased bases for transverse spatial light modes of dimension~7. Colour codings for intensity and phase are shown below in arbitrary units from 0 to 1 and 0 to $2\pi$, respectively.  a) Eigenstates of the generator $Z_7$ , which are also known vortex modes or OAM eigenstates (intensity null at the center of the beam due to the phase singularity is too small to be seen). b) Eigenstates of the $X_7$ operator can be described by so-called angle modes due to their intensity profil. c) Theoretical plot of intensity and phase of the eigenstates of the third mutually unbiased basis, which is constructed by $X_7Z_7$.}}
    \label{fig:3MUBs}
\end{figure}
With the promising results of an increased secret key rate at hand, we now turn to laboratory implementations of discrete high-dimensional state spaces. Although the presented theoretical analysis is valid for any experimental realization, we focus on one prominent example, namely transverse spatial light modes. Encoding high-dimensional quantum states on the orbital angular momentum of photons is a vibrant field in which technologies to generate and manipulate the states have matured over the last 15 years. Here, the eigenstates of two MUBs can be intuitively understood as the complementary variables, orbital angular momentum (OAM) and angular position (ANG). They correspond to the generators $Z_d$ and $X_d$, respectively, which we introduced earlier (Eq.~(\ref{eq:PauliGens})). High-dimensional states of both MUBs have been used in previous experiments to demonstrate complementarity as well as high-dimensionality of the generated quantum states~\cite{leach2010quantum,krenn2014generation,edgar2012imaging}. More importantly, their advantage in high-dimensional QKD has been demonstrated recently~\cite{Mirhosseini:2015fy} and experimental techniques for efficiently sorting the encoded qudits are well established \cite{Mirhosseini:2013em, Berkhout2010}.

In Fig.~\ref{fig:3MUBs}, we give an example of the eigenmodes of all three MUBs for dimension $d=7$: the OAM-basis $Z_7$, the ANG-basis $X_7$ and the eigenstates of $X_7Z_7$. The typical vortex of OAM carrying light modes and their according helical phase dependence (from which the OAM stems) can be seen (Fig.~\ref{fig:3MUBs}.~a) as well as the angular-shaped intensity of the states in the second MUBs (Fig.~\ref{fig:3MUBs}.~b). The modes of the third MUB are more complex in their intensity and phase profile (Fig.~\ref{fig:3MUBs}.~c), which leads to open questions of how practical such modes are in a laboratory setting. Although modern techniques to generate complex light fields with high fidelity and efficiency are well known \cite{arrizon2007pixelated}, the efficient sorting of a general set of spatial modes remains difficult. Possible techniques will need to be efficient and to work on the single photon level. Both requirements are fulfilled for established sorting devices that are used for OAM and ANG modes but no direct techniques is known yet, which sorts the modes of the third basis. One way to circumvent this lack of an efficient direct sorting would be to transfer the transverse spatial degree of freedom into different optical paths, e.g. as described~\cite{fickler2014interface}. Once transferred, it is known how to realize any unitary transformation on the state, and thus an efficient detection could be done in any basis~\cite{reck1994experimental}. Here, the fast progress in integrated quantum optics might a promising way to realize such a so-called multiport even for dimensions as high as $d=7$ \cite{carolan2015universal,schaeff2015experimental}.

In summary, we calculated secret key rates and tightly estimated achievable upper bounds on acceptable errors for an asymptotic and finite key length scenario in high-dimensional QKD schemes. We were able not only  to reproduce and streamline already known bounds but mainly we (i) adapted the uncertainty-relations-based method to high-dimensional QKD with two MUBs leaving us with nonzero secret key rates even for a relatively small number of signals and (ii) extended the findings to a QKD scheme involving 3 MUBs basis. Given the assured existence of 3 MUBs in any dimension, our results are not limited to dimensions where the exact number of MUBs is known and they can be readily applied to laboratory implementations. Additionally, we give an example for a possible physical implementation, transverse spatial modes, for which mature techniques in generating all possible qudit-states exist and devices to efficiently sort the states of two MUBs are established. Hence, an important future challenge is to develop a practical device that efficiently sorts the modes of the third MUB. Given the derived increase in the secret key rate, the development of such a novel sorter will further boost high-dimensional QKD schemes and their real-world implementations.

\section*{Acknowledgement}

RB, RF and KB thank the Canada Excellence Research Chairs program for support.  AB, RB, and KB acknowledge support from The Natural Sciences and Engineering Research Council of Canada.  RB and MM acknowledge support from the US Office of Naval Research. In addition, KB thanks Patrick Coles for comments and pointers to relevant literature (\cite{ferenczi2012symmetries,sheridan2010security,coles2015unstructured}).

\bibliographystyle{unsrt}

%\bibliography{QKDOAM}

\end{document}